\documentclass[aps,prl,twocolumn,showpacs,superscriptaddress]{revtex4-1}
\usepackage{graphicx}

\begin{document}

\title{Short-range correlations in the magnetic ground state of Na$_4$Ir$_3$O$_8$}

\author{Rebecca Dally}
\affiliation{Department of Physics, Boston College, Chestnut Hill, Massachusetts 02467, USA}
\author{Tom Hogan}
\affiliation{Department of Physics, Boston College, Chestnut Hill, Massachusetts 02467, USA}
\author{Alex Amato} 
\affiliation{Laboratory for Muon Spin Spectroscopy, Paul Scherrer Institut, CH-5232 Villigen PSI, Switzerland}
\author{Hubertus Luetkens}
\affiliation{Laboratory for Muon Spin Spectroscopy, Paul Scherrer Institut, CH-5232 Villigen PSI, Switzerland}
\author{Chris Baines}
\affiliation{Laboratory for Muon Spin Spectroscopy, Paul Scherrer Institut, CH-5232 Villigen PSI, Switzerland}
\author{Jose Rodriguez-Rivera}
\affiliation{NIST Center for Neutron Research, National Institute of Standards and Technology, Gaithersburg, Maryland 20899, USA}
\author{Michael J. Graf}
\affiliation{Department of Physics, Boston College, Chestnut Hill, Massachusetts 02467, USA}
\author{Stephen D. Wilson}
\email{stephendwilson@engineering.ucsb.edu}
\affiliation{Department of Materials, University of California, Santa Barbara, California 93106, USA.}

\begin{abstract}
The magnetic ground state of the $J_{eff}=1/2$ hyper-Kagome lattice in Na$_4$Ir$_3$O$_8$ is explored via combined bulk magnetization, muon spin relaxation, and neutron scattering measurements.  A short-range, frozen, state comprised of quasi-static moments develops below a characteristic temperature of $T_F$=6 K, revealing an inhomogeneous distribution of spins occupying the entirety of the sample volume.  Quasi-static, short-range, spin correlations persist until at least $20$ mK and differ substantially from the nominally dynamic response of a quantum spin liquid. Our data demonstrate that an inhomogeneous magnetic ground state arises in Na$_4$Ir$_3$O$_8$ driven either by disorder inherent to the creation of the hyper-Kagome lattice itself or stabilized via quantum fluctuations.   
\end{abstract}


\maketitle

Models of spin-orbit entangled $J_{eff}=1/2$ electrons on edge-sharing octahedra have shown that the symmetric portions of the magnetic Heisenberg exchange coupling may cancel \cite{jackelli, ashvin}. This opens the possibility for an antisymmetric, bond-dependent magnetic exchange that can be mapped into a Hamiltonian with a spin liquid ground state \cite{kitaev, ashvin}.  In two-dimensions, the honeycomb lattice of (Li,Na)$_2$IrO$_3$ has been proposed to be close to this spin liquid regime \cite{jackelli2, gegenwart}, and in three-dimensions the leading candidate for realizing this new spin-orbit driven spin liquid is the hyper-Kagome lattice of Ir moments in Na$_4$Ir$_3$O$_8$ (Na-438) \cite{takagi}.  

While the geometric frustration of Ir moments on the hyper-Kagome lattice, the theorized $J_{eff}=1/2$ ground state, and the edge-sharing octahedra in Na-438 comprise the theoretical requirements for stabilizing a spin liquid phase, a range of ordered magnetic ground states may instead stabilize depending on the relative strengths of competing exchange parameters as well as the relevance of Dzyaloshinskii-Moriya (DM) interactions \cite{balents}. As a result, a number of ordered states have also been proposed, ranging from fluctuation-driven nematic order \cite{ybkim} to a variety of antiferromagnetic states \cite{balents,ashvin}. Which magnetic ground state in Na-438 is realized, however, remains experimentally unresolved.  

A high degree of magnetic frustration in Na-438 was initially suggested via the measurement of a large Curie-Weiss temperature $\Theta_{CW}\approx680$ K and the absence of spin freezing above $6$ K \cite{takagi}.  Magnetic heat capacity data, while revealing an anomalous peak near $30$ K, similarly have shown no signature of ordering down to $0.5$ K along with a linear term in the low temperature $C_{mag}(T)$---suggestive of gapless spin excitations in the ground state \cite{takagi, gegenwart}.  Furthermore, measurements of the magnetic Gr{\"u}neisen parameter have hinted at a nearby quantum critical point \cite{gegenwart2}.  While these studies have suggested an exotic spin liquid in the $J_{eff}=1/2$ hyper-Kagome lattice, direct experimental probes of magnetic correlations in this material are notably absent.  Thus the key question remains:  In the ground state, do spins remain predominantly dynamic as expected for a spin liquid, or do spins ultimately freeze into static/quasi-static correlations?

Here we use muon spin relaxation ($\mu$SR), bulk magnetization, and neutron scattering studies to resolve that the hyper-Kagome lattice transitions into a quasi-static state frozen below $6$ K with slow dynamics persisting down to $20$ mK. Our $\mu$SR data reveal that an irreversibility in the static spin susceptibility below $6$ K arises from the bulk of the spins in the system freezing and not a dilute impurity effect. The resulting magnetic ground state differs qualitatively from that expected of an intrinsically dynamic quantum ground state such as a quantum spin liquid, and, instead, moments freeze into a configuration of densely packed spins with short-range correlations confined to the length scale of one unit cell.  Our combined results point toward an unusual ground state realized by $J_{eff}$=1/2 spins on a hyper-Kagome lattice with long-time scale, persistent fluctuations that coexist with spin freezing and resemble those observed in quantum spin ice and related materials.  The recovered spin entropy in this material \cite{gegenwart2} can therefore be attributed to the onset of a short-range ordered magnetic ground state.

Zero field and longitudinal field $\mu$SR measurements were made at the Paul Scherrer Institute in a gas-flow cryostat over the range $1.6$ K $< T < 13$ K and in a dilution refrigerator over the range $20$ mK $< T < 1.3$ K \cite{supplemental}. Data were analyzed using Musrfit software \cite{musrfit}; for a general review of the $\mu$SR technique, see Refs. \cite{Amato} and \cite{Blundell}.  Magnetization measurements were performed in a Quantum Design MPMS3 magnetometer, and neutron scattering measurements were performed on the MACS spectrometer at the NIST Center for Neutron Research.  

\begin{figure}
\includegraphics[scale=.16]{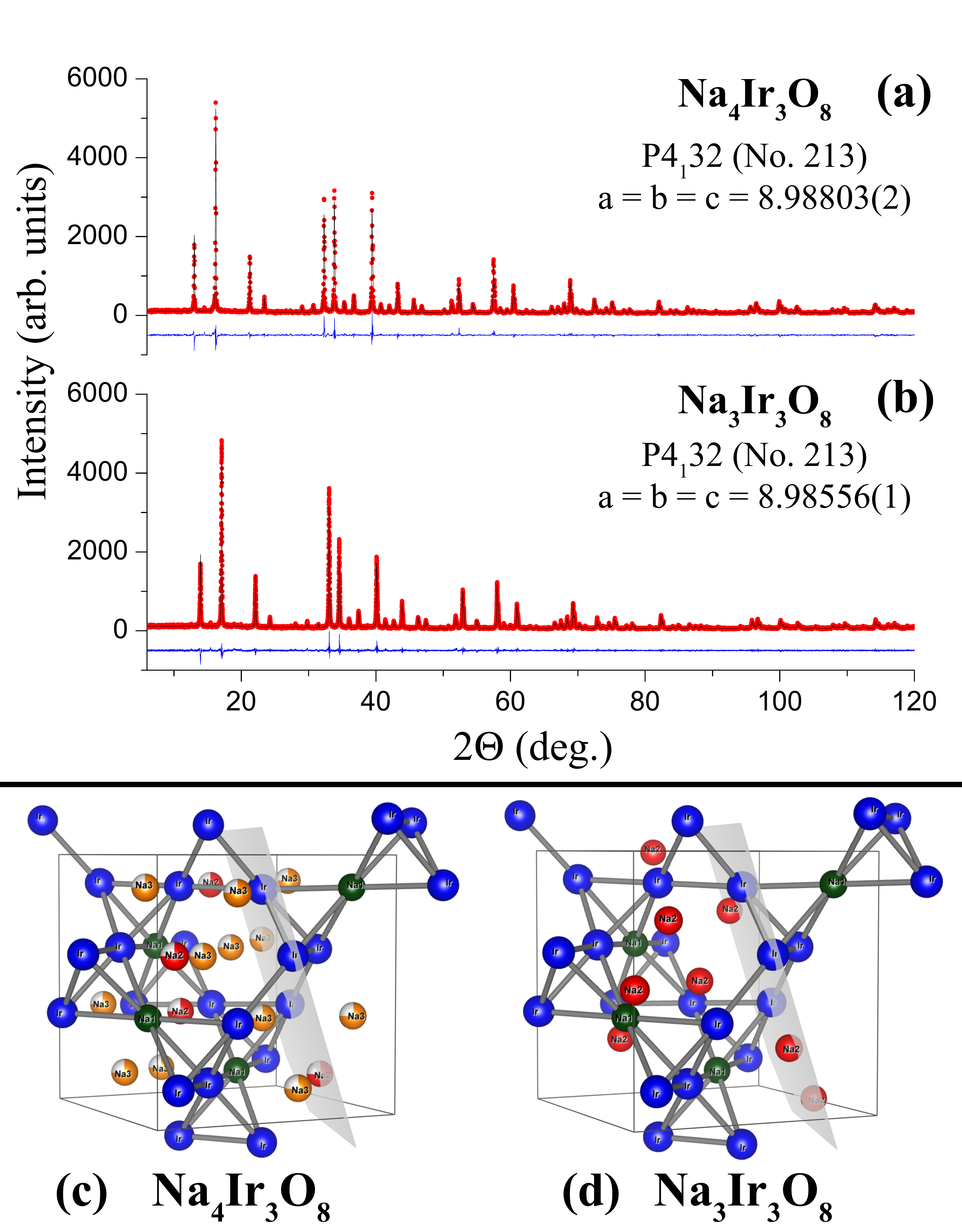}
\caption{Room temperature X-ray diffraction measurements collected with a Cu-K$_{\alpha}$ source for (a) Na-438 and (b) Na-338.  Panels (c) and (d) show the crystal structures of Na-438 and Na-338.  Ir atoms are denoted as blue spheres, and Na(1), Na(2), and Na(3) sites are denoted by green, red and orange spheres respectively. Oxygen atoms are not plotted for clarity, and the (3,1,1) plane is highlighted. }
\end{figure}
 
\begin{figure}
\includegraphics[scale=.6]{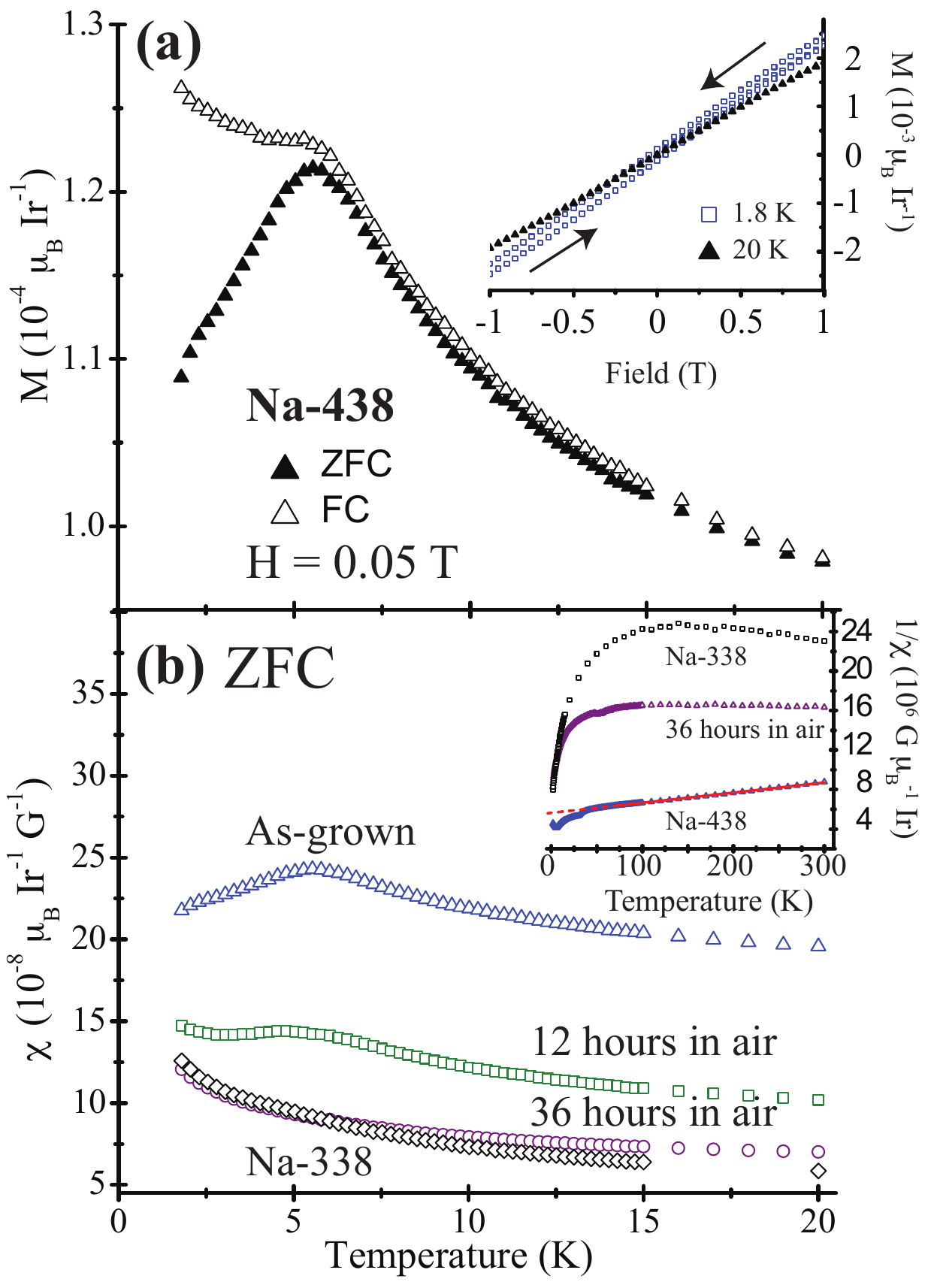}
\caption{(a) Bulk magnetization collected for Na-438 under both FC (closed symbols) and ZFC (open symbols).  The inset shows the magnetization, $M$, versus applied field $H$ both above and below $T_F$.  (b) ZFC static spin susceptibility plotted for as-grown Na-438 (blue triangles) as well as Na-438 left in air for 12 hours (green squares), 36 hours (purple circles), and converted Na-338 (black diamonds).  Inset shows $\chi^{-1}(T)$ for as-grown Na-438, 36 hour exposed Na-438, and Na-338.  Line is Curie-Weiss fit. }
\end{figure}

Polycrystalline Na$_4$Ir$_3$O$_8$ (Na-438) was prepared in a manner similar to earlier reports \cite{takagi,gegenwart2}, and polycrystalline Na$_3$Ir$_3$O$_8$ (Na-338) was prepared by converting Na-438 into Na-338 via post-reaction chemical treatment \cite{supplemental}.  Phase purity of each powder was confirmed via x-ray diffraction measurements shown in Fig. 1.  We found that Na-438 gradually decomposes upon exposure to atmosphere into Na$_{3+x}$Ir$_3$O$_8$ with a structure similar to the recently reported Na-338 structure \cite{na338}; hence care was taken to minimize its exposure to air and samples were manipulated in an inert atmosphere.  This behavior is similar to the decomposition reported in another high alkali content iridate, Na$_2$IrO$_3$ \cite{cava}.  The precise evolution of the structure of Na-438 once exposed to atmosphere is unknown, and we use the intentionally converted Na-338 as a reference for the electronic response of Na-deficient (ie. carrier doped) Na-438.  
  
Susceptibility measurements for both Na-438 and Na-338 are plotted in Fig. 2.  A Curie-Weiss fit for Na-438 ($75$ K$<T<300$ K),  shown in the inset of Fig. 2 (b) yielded a $\Theta_{CW}=555$ K and an effective local moment of $2.10$ $\mu_{B}$.  The local moment is larger than the maximum moment expected for a simple $S=\frac{1}{2}$ system, consistent with the larger values from earlier reports \cite{takagi, gegenwart2}. We stress here that this is an effective fit, and the high temperatures required for a quantitative fit are inaccessible due sample decomposition.  Below the breakdown of Curie-Weiss behavior, a cusp appears peaked at $T_F=6$ K in the zero-field cooled (ZFC) susceptibility. This peak coincides with the onset of irreversibility when compared with field-cooled (FC) data, and field sweeps at $1.8$ K reveal the onset of a frozen state with a finite coercive field (Fig. 2 (a) inset).  Once powder from this same batch is converted into Na-338, the nominal doping of $\frac{1}{3}$ of Ir-sites to Ir$^{5+}$ causes this cusp to disappear.

The frozen state below $6$ K in Na-438 is fragile and gradually diminishes upon exposure to air as shown in Fig. 2 (b), and the structural conversion can be tracked via x-ray measurements  \cite{supplemental}.  We note here that high-temperature $M(T)$ data of Na-338 exhibits a positive slope \cite{na338}, and hence any admixture of Na-338 within Na-438 will increase the apparent value for $\Theta_{CW}$ (Fig. 2 (b) inset). This may explain the variability in previous susceptibility measurements \cite{takagi, gegenwart2}.     
  
\begin{figure}
\includegraphics[scale=.4]{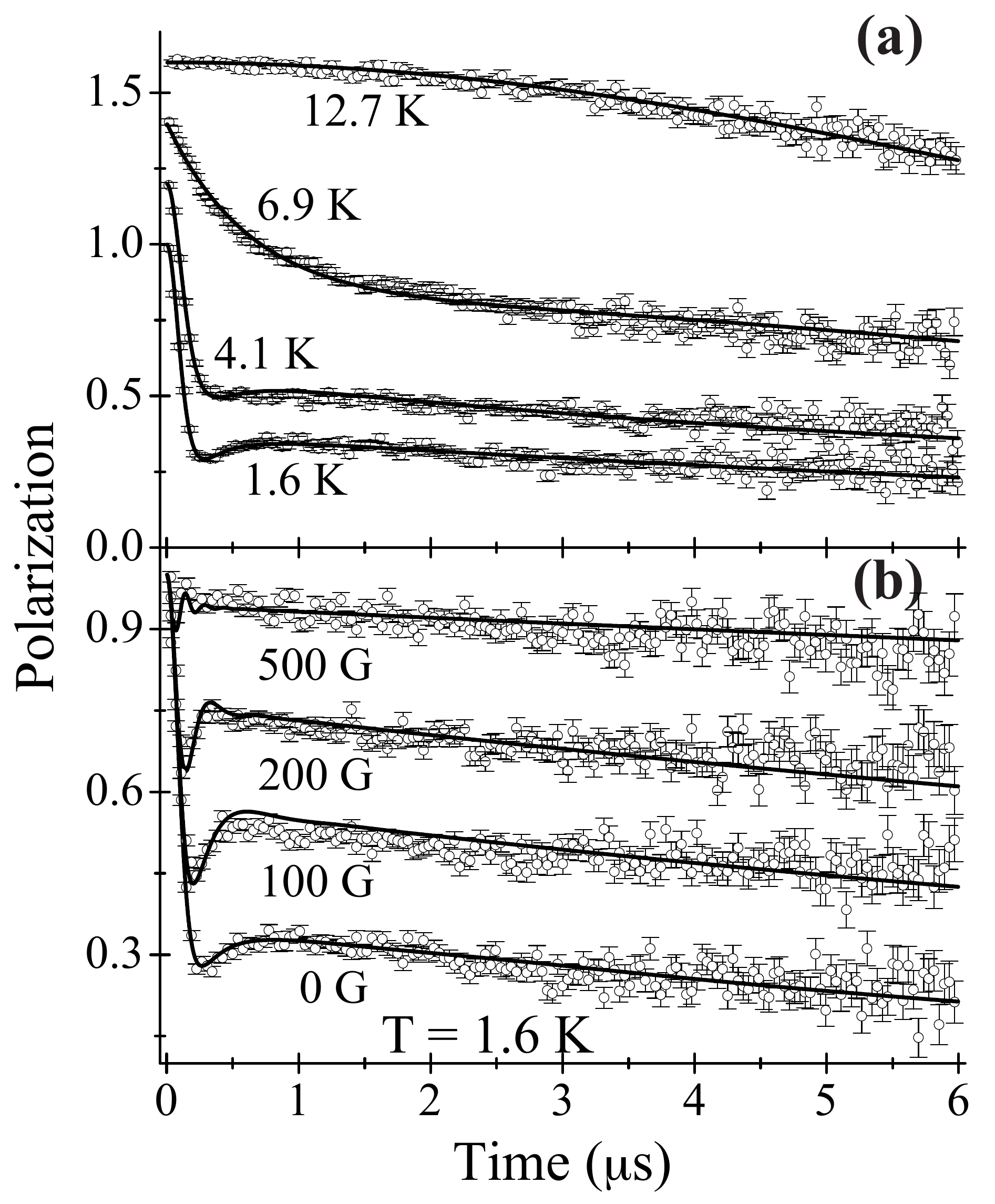}
\caption{(a) Time dependence of the muon polarization at several temperatures (open symbols), along with the fits (solid lines), as described in the text. The curves are offset by equal increments of 0.2 for clarity (b) Muon depolarization at 1.6 K as a function of time at several longitudinal field values. Open symbols are the data, and the solid lines are fits described in the text.  The error bars indicate one standard deviation.}
\end{figure}

In order to determine the origin of the irreversibility observed below $6$ K in Na-438, we performed $\mu$SR measurements. $\mu$SR is volumetric probe capable of differentiating between local spin freezing from dilute impurities and a global transition of spins into a state with quasi-static order.  Fig. 3 (a) shows the zero-field depolarization curves between $T=12.7$ K and $T=1.6$ K. The $12.7$ K curve shows a very slow depolarization, typical of muon depolarization via small nuclear moments of fixed size and random orientation. This can be fit by the Gaussian Kubo-Toyabe depolarization function \cite{Yaouanc}, 
\begin{equation}
G_{GKT}(t)=\frac{2}{3}(1-\Delta_N^2t^2)e^{\frac{-\Delta_N^2t^2}{2}}+\frac{1}{3}
\end{equation}
where $\Delta_N$ is proportional to the root-mean-square of the local field distribution (assumed to be Gaussian) due to the nuclei at the muon stopping site, $\Delta_N=\gamma_{\mu}\langle B_{loc}\rangle_{rms}$, and $\gamma_{\mu}$ is the muon gyromagnetic ratio. The ``$\frac{1}{3}$ tail'' in Eq. (1) is a defining characteristic of depolarization by random local fields that are static on the muon timescale.  From our fits at $12.7$ K , we find $\Delta_N=0.102(1)$ $\mu s^{-1}$, a value consistent with depolarization by nuclear moments.     

Below $10$ K the depolarization takes on an exponential form, and the rate increases sharply with decreasing temperature. As the sample is cooled below $6.5$ K, where the maximum in the static susceptibility is observed, we fit the depolarization with the phenomenological Gaussian-broadened Gaussian (GbG) function, $G_{GbG}(t)$, modified with a slow exponential decay \cite{Naokes1, Naokes2, Yaouanc}. The form for the polarization is:
\begin{equation}
G(t)=(1-f)G_{GbG}(t)e^{-\lambda t}+fe^{-\lambda_{338}t}
\end{equation}
and 
\begin{equation}
G_{GbG}(t)=\frac{2}{3}\left( \frac{1+R^2}{\alpha}\right)^{\frac{3}{2}}\left(1-\frac{\Delta_{W}^2t^2}{\alpha}\right)e^{\frac{-\Delta_{W}^2t^2}{2\alpha}}+\frac{1}{3}
\end{equation}

Here,  $\alpha=1+R^2+R^2\Delta_W^2t^2$, $\Delta_W^2=(W^2+\Delta_0^2)^{\frac{1}{2}}$, and $R = W / \Delta_0$.  This GbG form is simply a variation on the Gaussian Kubo-Toyabe function where the single local field distribution due to densely packed electronic moments is replaced by a Gaussian distribution of distributions with a mean value $\Delta_0$ and width $W$.  The form is known to describe systems where the muon depolarization is driven by disordered static magnetic moments with short-ranged correlations, e.g., the spin ice candidate Yb$_2$Ti$_2$O$_7$ \cite{yb227} and the doped skutterudite PrOs$_4$Sb$_{12}$ \cite{mac}.  A similar form of relaxation for muons in the presence of short-range order was recently derived analytically using statistical methods, providing a firm theoretical foundation for this association \cite{yaounanc2}.

We allowed for a refined volume fraction of a Na-338 impurity phase, $f$, which is characterized by slow exponential damping with a nearly temperature independent rate $\lambda_{338}$. This additional damping term $\lambda_{338}$ was independently measured \cite{supplemental}, and the temperature independent fraction $f$ was determined from the fits to the data at $1.6$ K.  The resulting fraction of impurity Na-338 phase within the nominally Na-438 powder was fit to be $6.6\%$, which likely arose from incidental exposure of the sample to atmosphere during transport.  The remaining $93.4\%$ of the sample volume below $6$ K is well described by the GbG formalism, confirming that the entirety of the sample transitions into a quasi-static state. 

In Fig. 4, we show the temperature variation of $\Delta_0$, $R$, and $\lambda$. The sudden increase in $\Delta_0$ and drop in $R$ coincide with the peak in the magnetic susceptibility. We note here that the mismatch in the data between $1.3$ and $1.6$ K is due to the presence of a large background component in the data taken in the dilution refrigerator due to the silver sample holder \cite{supplemental}.  Both $\Delta_0$ and $\lambda$ parameters approach constant values upon cooling in contrast to the behavior of a canonical spin glass. Based on the more accurate data taken in the gas flow cryostat, we take the low temperature values of $\Delta_0$, $R$, and $\lambda$ to be approximately $6.0$ MHz, $0.4$ and $0.1$ $\mu s^{-1}$, respectively. From this value of $\Delta_0$ we extract a characteristic field  $\langle B\rangle _{rms} = \Delta_0/\gamma_{\mu} = 70$ G. The muon stopping site is unknown, but assuming that the muon experiences a dipolar field due to an Ir$^{4+}$ magnetic moment located about half a unit cell away ($a = 8.988$ \AA), we roughly estimate $\mu_{Ir} \approx 0.5$ $\mu_{B}$, reasonable for the $J_{eff} = \frac{1}{2}$ state.

As an additional check of the quasi-static nature of the moments in Na-438, we also carried out measurements mapping the effect of a longitudinal magnetic field on the muon depolarization at $T = 1.6$ K. The data plotted in Fig. 3 (b) show that the $\frac{1}{3}$ tail at long time scales begins to be removed in an applied field of $H\gtrsim \langle B\rangle _{rms}$ and is completely suppressed under $\approx 500$ G.  This is an independent confirmation of the quasi-static nature of the depolarization and highlights the qualitative departure of the spin response from the dynamical, nearly field-independent, depolarization behavior expected for a spin liquid that only partially freezes below $T_F$ \cite{uemura1}.  

We approximated the longitudinal field dependence of the GbG function by a finite sum of GKT functions shown in Fig. 3 (b) that each include fluctuations and an applied longitudinal field \cite{Amato, Blundell}. Fits to this form under zero field yielded $\lambda$, $\Delta_{0}$, and $R$ values consistent with those of zero-field GbG fits.  Changing the magnetic field values while keeping these parameters fixed yields good agreement with the data under applied fields, and validates our conclusion that the muon depolarization is caused by a quasi-static magnetic field distribution dressed with slow fluctuations.

\begin{figure}
\includegraphics[scale=0.325]{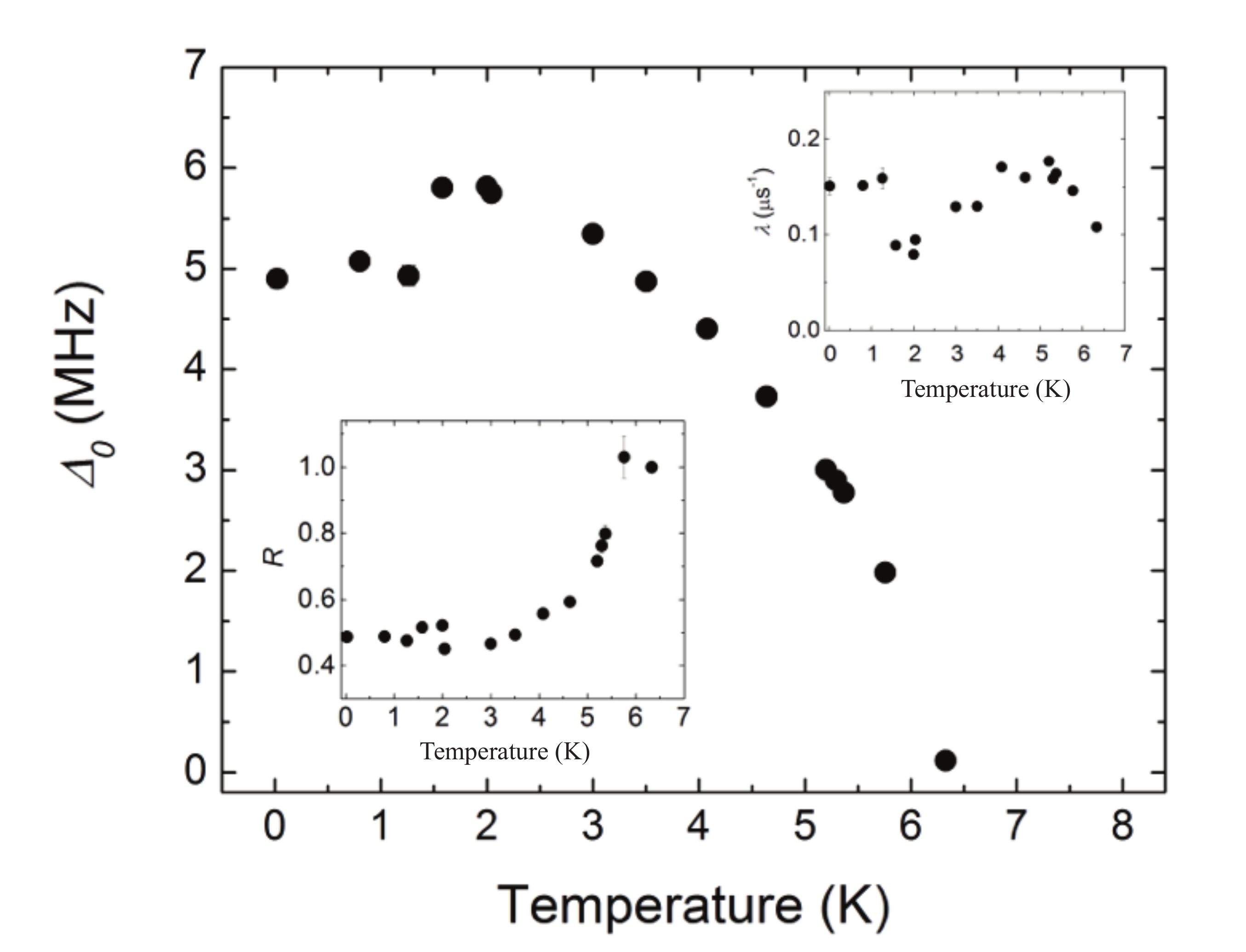}
\caption{The mean value of the Gaussian-broadened Gaussian distribution as a function of temperature. Upper inset: variation of the relaxation rate with temperature. Lower inset: the temperature dependence of the ratio of width of distributions to the distribution mean.  The error bars indicate one standard deviation of the fit parameters.}
\end{figure}

Monte Carlo simulations \cite{Naokes2} have shown that the ratio $R = W / \Delta_0$ is inversely proportional to the magnetic correlation length. In Na-438, the saturated region below $T_F$ shows a fit $R=0.4$, implying correlations that extend only over several nearest neighbor distances.  Freezing of moments into a quasi-static state below $6$ K can preclude the formation of a purely dynamical quantum spin-liquid ground state in Na-438.  Instead, quasi-static moments coexist with long-time scale spin fluctuations, roughly mirroring the coexisting dynamic and freezing behavior reported in geometrically frustrated magnets such as $S=1$ NiGa$_2$S$_4$ \cite{mclaughlin1} and $S=\frac{1}{2}$ Cu$_3$Ba(VO$_5$H)$_2$ \cite{wills}.  The degree of coexistence in this intermediate regime depends on the details of each system varying from predominantly static moments in Na-438 to the predominantly dynamic liquid state in Cu$_3$Ba(VO$_5$H)$_2$; however it appears to be a generic feature of a proximate, highly degenerate, quantum ground state.  

It is worth noting that systems far from frustration and well into an ordered regime may also show inhomogeneous short-range order and persistent fluctuations coexisting with conventional static order \cite{mclaughlin2,disseler}.  The small $\frac{T_F}{\Theta_{CW}}$ ratio in Na-438 however suggests that the dichotomy between quasi-static order and persistent fluctuations is driven via interactions characteristic of frustration. This combined with the anomalous thermodynamic properties of Na-438 \cite{gegenwart2, takagi} parallels the configurationally degenerate phases with fluctuating order (CDFO) observed in spin-ice and related phases \cite{gardner, yb227, king} which possess short-range correlations that continue to fluctuate as $T\rightarrow 0$.  

Theoretical studies have shown that an unfrustrated classical ground state should stabilize in Na-438 in the limit of strong spin-orbit interactions and direct Ir-O-Ir superexchange; however for direct Ir-Ir exchange, quantum fluctuations may remain relevant depending on the details of DM effects \cite{balents}.  A Kitaev-Heisenberg model of Na$_4$Ir$_3$O$_8$ has similarly predicted an array of classically ordered phases with select regions of parameter space allowing for quantum disordered ground states  \cite{ashvin}. A nematic order parameter, driven by fluctuations, has also been predicted to stabilize in Na-438 via an order-by-disorder mechanism \cite{ybkim}. Determining precisely which of these spin configurations freezes requires a momentum resolved probe such as neutron scattering.  Our initial neutron scattering data reveal no sign of magnetic correlations  \cite{supplemental}---implying a small ordered moment difficult to resolve in powder measurements.  Future measurements on single crystals will be required to fully characterize the frozen state; however our data does allow models predicting the component field distribution felt by the muon to be tested using the approach of Yaouanc et al. \cite{yaounanc2}.    

Our observations are consistent with the formation of a CDFO phase in Na-438; however the freezing transition at $6$ K is curiously absent from previously reported heat capacity data. Instead heat capacity reveals only a broad peak at $30$ K that is also only weakly field dependent, presenting a theoretical challenge in reconciling the details of spin freezing and the higher temperature entropy release in this material.   At lower temperatures, however, our data show that the magnetic entropy observed arises not from spins possessing purely dynamic, liquid-like, correlations in the quantum ground state but rather from frozen spins likely belonging to a growing class of CDFO phases that exhibit long-time scale damping coexisting within a short-range ordered, quasi-static spin state.
 
\acknowledgments{
We would like to thank Sean Giblin and Ram Seshadri for preliminary SQUID measurements.  This work was supported in part by NSF CAREER award DMR-1056625 (S.D.W.). Muon experiments were performed at the Swiss Muon Source at the Paul Scherrer Institute (Switzerland).  This work utilized facilities supported in part under NSF award DMR-0944772 and SQUID measurements were supported in part by grant DMR-1337567}


\begin{thebibliography}{}
\bibitem{jackelli} G. Jackeli and G. Khaliullin, Phys. Rev. Lett. 102, 017205 (2009).
\bibitem{ashvin} Itamar Kimchi and Ashvin Vishwanath, Phys. Rev. B. 89, 014414 (2014).
\bibitem{kitaev} A. Kitaev, Ann. Phys. (N.Y.) 321, 2 (2006).
\bibitem{gegenwart} Yogesh Singh and P. Gegenwart, Phys. Rev. B 82, 064412 (2010).
\bibitem{jackelli2} Jiri Chaloupka, George Jackeli, and Giniyat Khaliullin, Phys. Rev. Lett. 110, 097204 (2013).
\bibitem{takagi} Yoshihiko Okamoto, Minoru Nohara, Hiroko Aruga-Katori, and Hidenori Takagi, Phys. Rev. Lett. 99, 137207 (2007).
\bibitem{balents} Gang Chen and Leon Balents, Phys. Rev. B 78, 094403 (2008).
\bibitem{ybkim} John M. Hopkinson, Sergei V. Isakov, Hae-Young Kee, and Yong Baek Kim, Phys. Rev. Lett. 99, 037201 (2007).
\bibitem{gegenwart2} Yogesh Singh, Y. Tokiwa, J. Dong, and P. Gegenwart, Phys. Rev. B 88, 220413 (2013).
\bibitem{uemura1} Y.J. Uemura, A. Keren, K. Kojima, L.P. Le, G. M. Luke, W. D. Wu, Y. Ajiro, T. Asano, Y. Kuriyama, M. Mekata, H. Kikuchi, and K. Kakurai, Phys. Rev. Lett. 73, 3306 (1994).
\bibitem{musrfit} A. Suter and B. M. Wojek, Phys. Procedia 30, 69 (2012).
\bibitem{na338} T. Takayama, A. Matsumoto, J. Nuss, A. Yaresko, K. Ishii, M. Yoshida, J. Mizuki, and H. Takagi, arXiv:1311.2885.
\bibitem{Amato} A. Amato, Rev. Mod. Phys. 69, 1119 (1997).
\bibitem{Yaouanc} A. Yaouanc and P. Dalmas de  R{\'e}otier, Muon Spin Rotation, Relaxation, and Resonance: Applications to Condensed Matter (Oxford University Press, New York, 2011).
\bibitem{Blundell} S. J. Blundell, Cont. Physics 40, 175 (1999).
\bibitem{supplemental} See Supplemental Information for further details.
\bibitem{cava} J. Krizan, J. H. Roudebush, G. M. Fox and R.J. Cava, Materials Research Bulletin 52, 162 (2014)
\bibitem{Naokes1} D. R. Noakes and G. M. Kalvius, Phys. Rev. B 56, 2352 (1997).
\bibitem{Naokes2} D. R. Noakes, J. Phys.: Condens. Matter 11, 1589 (1999).
\bibitem{yb227} J. A. Hodges, P. Bonville, A. Forget, A. Yaouanc, P. Dalmas de R{\'e}otier, G. Andr{\'e}, M. Rams, K. Kr{\'o}las, C. Ritter, P. C. M. Gubbens, C. T. Kaiser, P. J. C. King, and C. Baines, Phys. Rev. Lett. 88, 077204 (2002).
\bibitem{mac} D. E. MacLaughlin, P.-C. Ho, L. Shu, O. O. Bernal, S. Zhao, A. A. Dooraghi, T. Yanagisawa, M. B. Maple, and R. H. Fukuda, Phys. Rev. B 89, 144419 (2014).
\bibitem{yaounanc2} A. Yaouanc, A. Maisuradze, P. Dalmas de  R{\'e}otier, Phys. Rev. B 87, 134405 (2013).
\bibitem{wills} R. H. Colman, F. Bert, D. Boldrin, A. D. Hillier, P. Manuel, P. Mendels, and A. S. Wills, Phys. Rev. B 83, 180416 (2011).
\bibitem{mclaughlin1}D. E. MacLaughlin, Y. Nambu, S. Nakatsuji, R. H. Heffner, Lei Shu, O. O. Bernal, and K. Ishida, Phys. Rev. B 78, 220403 (2008).
\bibitem{mclaughlin2} Songrui Zhao, J. M. Mackie, D. E. MacLaughlin, O. O. Bernal, J. J. Ishikawa, Y. Ohta, and S. Nakatsuji, Phys. Rev. B 83, 180402 (2011).
\bibitem{disseler} S. M. Disseler, Chetan Dhital, A. Amato, S. R. Giblin, Clarina de la Cruz, Stephen D. Wilson, and M. J. Graf, Phys. Rev. B 86, 014428 (2012).
\bibitem{gardner} J. S. Gardner, S. R. Dunsiger, B. D. Gaulin, M. J. P. Gingras, J. E. Greedan, R. F. Kiefl, M. D. Lumsden, W. A. MacFarlane, N. P. Raju, J. E. Sonier, I. Swainson, and Z. Tun, Phys. Rev. Lett. 82, 1012 (1999).
\bibitem{king} P. Dalmas de R{\'e}otier, A. Yaouanc, L. Keller, A. Cervellino, B. Roessli, C. Baines, A. Forget, C. Vaju, P. C. M. Gubbens, A. Amato, and P. J. C. King, Phys. Rev. Lett. 96, 127202 (2006).
\end{thebibliography}

\end{document}